\documentclass[sigconf,anonymous=False]{acmart}

\AtBeginDocument{%
	\providecommand\BibTeX{{%
			\normalfont B\kern-0.5em{\scshape i\kern-0.25em b}\kern-0.6em\TeX}}}

\usepackage{subcaption}
\usepackage{graphicx}
\usepackage{caption}
\usepackage{xcolor}
\usepackage{amsmath}
\usepackage{titlesec}

\titlespacing*{\section}{0pt}{6pt plus 2pt}{6pt minus 2pt}
\titlespacing*{\subsection}{0pt}{6pt plus 2pt}{6pt minus 2pt}
\titlespacing*{\subsubsection}{0pt}{6pt plus 2pt}{6pt minus 2pt}
\copyrightyear{2021}
\acmYear{2021}
\setcopyright{acmcopyright}\acmConference[ICTIR '21]{Proceedings of the 2021 ACM SIGIR International Conference on the Theory of Information Retrieval}{July 11, 2021}{Virtual Event, Canada}
\acmBooktitle{Proceedings of the 2021 ACM SIGIR International Conference on the Theory of Information Retrieval (ICTIR '21), July 11, 2021, Virtual Event, Canada}
\acmPrice{15.00}
\acmDOI{10.1145/3471158.3472258}
\acmISBN{978-1-4503-8611-1/21/07}


\begin{document}
\fancyhead{}

\title{Understanding the Effectiveness of Reviews in E-commerce Top-N Recommendation}

\author{Zhichao Xu}
\email{zhichao.xu@utah.edu}
\affiliation{%
	\institution{University of Utah}
}

\author{Hansi Zeng}
\email{hanszeng@cs.utah.edu}
\affiliation{%
\institution{University of Utah}
}

\author{Qingyao Ai}
\email{aiqy@cs.utah.edu}
\affiliation{%
	\institution{University of Utah}
}

\begin{abstract}
%There are two major tasks in the study of recommender systems, rating prediction and top-N recommendation. 
%In rating prediction, an user's affinity towards an item is expressed explicitly through a numerical rating. Thus it is often formulated as an explicit feedback problem.
% Rating prediction aims to accurately predict the rating given by the user to the item and is often formulated as an explicit feedback problem. 
%In top-N recommendation, the user's affinity towards the item is expressed implicitly through certain kinds of behaviors, such as view, add-to-cart or purchase. 
% Top-N recommendation aims to predict the interact-or-not behaviors between the user and the candidate items. 
%Thus, the task is often formulated as an implicit feedback problem. 
Modern E-commerce websites contain heterogeneous sources of information, such as numerical ratings, textual reviews and images. These information can be utilized to assist recommendation. Through textual reviews, a user explicitly express her affinity towards the item. 
Previous researchers found that by using the information extracted from these reviews, we can better profile the users' explicit preferences as well as the item features, leading to the improvement of recommendation performance. However, most of the previous algorithms were only utilizing the review information for explicit-feedback problem i.e. rating prediction, and when it comes to implicit-feedback ranking problem such as top-N recommendation, the usage of review information has not been fully explored.
Seeing this gap, in this work, we investigate the effectiveness of textual review information for top-N recommendation under E-commerce settings. 
We adapt several SOTA review-based rating prediction models for top-N recommendation tasks and compare them to existing top-N recommendation models from both performance and efficiency.
We find that models utilizing only review information can not achieve better performances than vanilla implicit-feedback matrix factorization method. 
When utilizing review information as a regularizer or auxiliary information, the performance of implicit-feedback matrix factorization method can be further improved. 
However, the optimal model structure to utilize textual reviews for E-commerce top-N recommendation is yet to be determined.
\end{abstract}


\maketitle
%!TEX root=CIKM2020_WEU_recsys.tex

\section{Introduction}
% Modern E-commerce websites contain heterogeneous sources of information. For instance, on Amazon, a user can leave a rating to an item accompanied with textual reviews and multiple images. Through these information, the user explicitly express her degree of likeness to the item. On the other hand, different from the some rating application, i.e. BookReviews or IMDB, on E-commerce website, the user needs to purchase the item before leaving a review. This purchase-or-not decision making process can be regarded as a implicit feedback problem. Rating prediction and top-N recommendation are two most important lines of works for explicit feedback and implicit feedback, respectively. 

There are two major lines of work in the study of recommender systems, rating prediction and top-N recommendation. 
In several recommendation applications, a user will give a numerical rating to the item after the interaction, such as 1-10 star ratings at the movie rating website IMDB. Through the numerical rating score, the user's level of affinity is expressed explicitly, so it is often formulated as an explicitly feedback problem. The task of rating prediction aims to accurately predict this numerical rating.
Top-N recommendation problem, on the other hand, focus on predicting what items the user is likely to interact with. The user implicitly express her affinity towards the item by making the interact-or-not decision, so it is often formulated as an implicit feedback problem. 
Different from some rating applications, i.e. IMDB, in modern E-commerce websites, the user needs to purchase the item before leaving a rating accompanied with textual reviews or images. 
This first-purchase-then-review nature makes E-commerce top-N recommendation challenging since many other factors i.e. the price of the item, are taken into consideration when making the purchase decision. 

A good news is that modern E-commerce websites contain heterogeneous sources of information, i.e. numerical ratings, textual reviews, images, which can be utilized to help with recommendation.
Through textual reviews, a user explicitly express her affinity towards the item. In both rating prediction and top-N recommendation, review data has been widely identified to be useful in improving recommendation performance. 
Ganu et al. \cite{ganu2009beyond} found that text information can be used to assist the recommendation procedure. 
%Wang et al. \cite{wang2011collaborative} introduced the topic model to study the textual reviews.
%McAuley et al. \cite{mcauley2013hidden} utilized the topic models to regularize the latent factors learned from rating matrix. They also introduced a large scale recommendation dataset which contains high quality review texts. 
Later, researchers \cite{ling2014ratings, diao2014jointly,zhang2017joint,chen2018neural,bao2014topicmf,ai2018learning,chen2016learning,xu2020commerce,zeng2021zero,zeng2020hierarchical,guan2019attentive,zhang2014explicit,he2015trirank,pan2020review} kept making progress in effectively incorporating review information into the task of recommendation. 
However, as pointed out by Peña et al. \cite{pena2020combining}, the exploration to utilize textual reviews for recommendation are limited to rating prediction, and the effectiveness of review-based models for top-N recommendation is still underexplored. This gap serves as our motivation for our work.

% In this paper, we study existing review-based explicit feedback model designs to see whether they can also be effective for implicit feedback top-N recommendation problem.
In this work, we focus on the discussion of the better model designs to utilize textual reviews for E-commerce top-N recommendation.
We replicate several state-of-the-art review-based rating prediction algorithms to make them suitable for top-N recommendation. 
% We make reasonable \& necessary modifications to the original algorithms, which includes adjusting the objective functions and the experimental settings. 
Furthermore, we also include existing top-N recommendation models and draw comparison from both performance and inference speed.
Our experiment results show that while reviews can be effective in the rating prediction task where users explicitly express their preferences, algorithms purely relying on features extracted from review data are outperformed by vanilla implicit feedback matrix factorization method.
On the other hand, using review information as an additional regularizer or as auxiliary information can further improve matrix factorization's performance.
Also, review-based models with complex neural structures are of high time complexity and are not suitable for online top-N recommendation in practice.
The optimal model structure to utilize textual reviews for E-commerce top-N recommendation is yet to be determined.

\section{Related Work}
In both item-rating applications i.e. IMDB and E-commerce websites i.e. Amazon, users explicitly express their affinities towards items by giving numerical scores and leaving textual reviews. 
Existing exploration in utilizing textual reviews for recommendation can be roughly classified into two categories: (1) text-as-feature: this line of research focused on utilizing textual information as the an information source to build user/item representations and conduct recommendations. Ganu \cite{ganu2009beyond} et al. propose to extract aspects from textual reviews to assist with rating prediction. Wang et al. \cite{wang2011collaborative} trained a topic model from content of scientific articles and combine it with the rating matrix to recommend. EFM \cite{zhang2014explicit} conducted aspect-level sentiment analysis to extract user's preference and product's quality on specific product features, then incorporate the results into matrix factorization framework to recommend. TriRank \cite{he2015trirank} modeled the user-aspect-item relations through a tripartite graph and cast the recommendation task to vertex ranking. 
DeepCoNN \cite{zheng2017joint} first utilizes a deep neural network to learn user and item representations separately and calculate the recommendation score using a similarity function. Later, researchers \cite{guan2019attentive,catherine2017transnets,chen2018neural,zeng2020hierarchical,zeng2021zero} applied more complex neural architectures to learn information from textual reviews and further improve the recommendation performances.
% AAFM \cite{guan2019attentive} utilized an attention network to learn the weights between different aspects of users and items.
(2) text-as-regularizers: this line of research focused on utilizing textual information to regularize the user-item interactions and conduct recommendations. fLDA \cite{agarwal2010flda}, extended from
matrix factorization, regularizes both user and item factors
in a supervised manner through explicit user features and
the bag of words associated with each item. HFT \cite{mcauley2013hidden} trains a topic model to regularize the latent factor model learned from the rating matrix. 
JRL utilizes the paragraph vector model \cite{le2014distributed} trained from textual reviews to regularizes the latent factor model trained from interaction matrix. To the best of our knowledge, it is the only model in this category that is originally designed for top-N recommendation task.

In this work, we focus on investigating the existing review-based text-as-feature models' performance in top-N recommendation. The closest work to ours is from Pena et al. \cite{pena2020combining}. They pointed out the exploration to utilize textual reviews for top-N recommendation is limited. Instead of proposing a new model to utilize textual reviews, we focus on studying existing works from both performance and efficiency. We hope to get insights from the results and inspire the community.
% \vspace{-1pt}
\section{Methodology}
\subsection{Problem Formulation}

We give a summary of notations used in this section in Table \ref{tab:notations}.
Let $|\mathcal{U}|$ and $|\mathcal{I}|$ denote the number of users and items, respectively. We define the user-item rating matrix from users' explicit feedback as $\mathcal{R}^{|\mathcal{U}| \times |\mathcal{I}|}$ where each interaction tuple $(u, i)$ is associated with an numerical rating $r_u^i$ and a piece of textual review $\gamma_u^i$ explaining why the user $u$ gave such rating $r_u^i$ to item $i$. 
Rating prediction task aims to minimize the pointwise MSE loss between $\hat{r}_u^i$ and $r_u^i$:
\begin{equation}
    \frac{1}{\mathcal{Y}} \sum_{(u,i) \in \mathcal{Y}} ||\hat{r}_u^i-r_u^i||^2
\end{equation}
where $\mathcal{Y}$ is the set of $(u,i)$ pairs that user $u$ has rated item $i$.

In contrast, Top-N recommendation task aims to provide each user with a set of \textit{N} items from a large set of items. Rendle et al. \cite{rendle2012bpr} proposed pairwise loss function (BPR) and has been widely used in top-N Recommendation from implicit feedback. 
Let $\mathcal{Y}^-$ be the set of negative $(u,i)$ pairs (e.g., if we consider user purchase as ground truth, then $\mathcal{Y}^-$ is the unpurchased $(u,i)$ pairs).
Given $\mathcal{Y}$ and $\mathcal{Y}^-$, the pairwise binary cross-entropy loss of BPR can be formulated as:
\begin{equation}
\begin{split}
    \mathcal{L} =  \!\!\!\!\!\!\!\!\!\!\!\!\!\!\!\sum_{(u,i) \in \mathcal{Y}, (u,j)\sim \mathcal{Y}^-}\!\!\!\!\! \!\!\!\!\!\!\!\!\!P_u(i\!>\!j)\log (P_u(i\!>\!j))\! +\! (1\!-\!P_u(i\!>\!j))\!\log (1\!-\!P_u(i\!>\!j)) 
\end{split}
\label{equ:loss}
\end{equation}
where $(u,j)$ is randomly sampled from $\mathcal{Y}^-$ based on $(u,i)$, and $P_u(i\!>\!j)$ is defined as
$$
P_u(i\!>\!j) = \frac{1}{1+\exp(y_u^i-y_u^j)}
$$
In our work, we modify the original algorithms designed for rating prediction by changing their loss functions from pointwise loss to pairwise loss for top-N recommendation. We also implement the corresponding negative sampling strategy for efficient training.

\begin{table}[t]
\caption{A summary of notations}
\begin{tabular}{l|l}
\hline
\hline
$u$, $\mathcal{U}$ & user, user set\\
$i$, $\mathcal{I}$ & item, item set\\
$r_u^i$ & ground-truth rating score of user-item pair $(u,i)$\\
$\hat{r}_u^i$ & predicted rating score of user-item pair $(u,i)$\\
$\gamma_u^i$ & textual review $\gamma$ given by $u$ to $i$ \\
$\Gamma_u$ & reviews set given by user $u$ \\
$\Gamma_i$ & reviews set given to item $i$ \\
$\beta_i$ & item bias \\ 
$\overrightarrow{u}$ & user latent factor \\
$\overrightarrow{i}$ & item latent factor \\
$\mathcal{Y}$ & positive user-item pairs set\\
$\mathcal{Y}^-$ & negative user-item pairs set \\
$y_u^i$ & predicted ranking score of user-item pair $(u,i)$\\
%$\hat{y}_u^i$ & predicted ranking score of user-item pair $(u,i)$\\
\hline 
\hline
\end{tabular}
\label{tab:notations}
\end{table}

\subsection{Recommendation Models}
To investigate the effectiveness of explicit review information in the task of top-N recommendation, we include a variety of state-of-the-art review-based models. We also consider to compare the review-based models with some classical implicit-feedback models for better evaluation. The models are classified into three different categories: the implicit-feedback interaction-based models, models using text information as feature, and models using text information as regularizer. We make necessary adjustments to adapt these models for top-N recommendation. The list of models we used in experiments is as follows:

\subsubsection{Interaction-based Models}
Interaction-based models simply model the historical interactions between users and items. They utilize a latent factor model, where each dimension of the latent factor is designed to represent a specific feature of the users and the items. Then a similarity function (mostly inner product) is applied to calculate the similarity between the user latent factor and the item latent factor and get the affinity score.

\textbf{Bayesian Personalized Ranking Matrix Factorization (BPR-MF)~\cite{koren2009matrix,rendle2012bpr}} BPR-MF follows the vanilla matrix factorization setup where each user \& item is represented by a latent factor. We use the pairwise loss from implicit feedback to train the model, and the final affinity score given by user $u$ to item $i$ is predicted as 
\begin{equation}
    y_u^i = \beta_i + \overrightarrow{u} \cdot \overrightarrow{i}
\end{equation}
and we optimize the parameters by maximize Equation \ref{equ:loss}.

\textbf{BPR Generalized Matrix Factorization (BPR-GMF)~\cite{he2017neural}} BPR-GMF adds a generalized function to model the complex interactions between user and item latent factors. Specifically, the affinity score is 
\begin{equation}
    y_u^i = \beta_i + \textit{F} (\overrightarrow{u} \cdot \overrightarrow{i})
    \label{ref:BPR-GMF}
\end{equation}
where $F$ is a deep neural network structure. In our implementation, we use a multi-layer densely-connected neural network (MLP) and we tune the number of MLP layers for best performance.

\subsubsection{Text-as-regularizer Models} \hfill \\
Text-as-regularizer models follows a traditional matrix factorization setup, and apply an additional objective function utilizing the representations learned from textual information to regularize the latent factors. The model can be trained offline and in the inference stage, the affinity score is computed by a simple inner product function. Thus they are considered an effective way for online recommendation.

\textbf{Hidden Factors and Topics (BPR-HFT)~\cite{mcauley2013hidden}} In addition to the BPR loss, BPR-HFT utilizes textual reviews to train an latent dirichlet allocation topic model and minimize the corpus likelihood to regularize the latent factors used in matrix factorization. 

\textbf{Joint Representation Learning (JRL)~\cite{zhang2017joint}} Motivated by the paragraph vector model \cite{le2014distributed}, JRL utilizes an additional generative loss built from textual reviews to regularize the latent vectors learned from implicit-feedback interaction matrix.

\subsubsection{Text-as-feature Models} 
Text-as-feature models utilize representations learned from textual reviews to build user/item feature vectors.
A similarity function is then applied to calculate the affinity scores.

\textbf{BERT-Rep \cite{nogueira2019bert}:} For each user, all her reviews are aggregated to form a long document and input to BERT \cite{devlin2018bert} to encode, and the output layer's [CLS] token vector is used as the user representation. The same procedure is applied to get the item representation. Then we apply dot product over the user/item representation to predict the corresponding affinity score as
\begin{equation}
    BERT(Rep)(u,i)=\overrightarrow{u}_{cls}^{last} \ast \overrightarrow{i}_{cls}^{last}
\end{equation}
where $\ast$ denotes inner product and $last$ denotes the last layer of BERT's Transformer network.

\textbf{Deep Co-operative Neural Network (DeepCoNN)~\cite{zheng2017joint}}: Different from previous algorithms \cite{mcauley2013hidden,bao2014topicmf,diao2014jointly}, DeepCoNN utilizes a CNN-based neural architecture to extract information from textual reviews. Specifically, all the reviews in $\Gamma_u$ are concatenated as a document, then a TextCNN architecture \cite{zhang2015sensitivity} is applied to extract the latent feature factors from review documents to form the user feature factor. The same procedure is applied to get the item feature factor.

\textbf{Neural Attentive Rating Regression (NARRE)~\cite{chen2018neural}}: Also based on TextCNN, NARRE additionaly learns a review-level attention weights distribution of each single piece of review in user/item document and it achieves better performance than DeepCoNN in terms of rating prediction.

\textbf{Multi-Pointer Co-Attention Network (MPCN) \cite{tay2018multi}}: MPCN selects informative reviews from $\Gamma_u$ \& $\Gamma_i$ by review-level pointers using co-attention technique, and selects informative word-level representations by applying word-level pointers over selected reviews.

\textbf{Asymmetrical Hierarchical Networks with Attentive Interactions (AHN) \cite{dong2020asymmetrical}}: AHN treats user and item asymmetrically and builds representations hierarchically from sentence level and review level. It also dynamically models the interaction using co-attention mechanism.

\textbf{Interpretable Convolutional Neural Networks with Dual Local and Global Attention (Dual-ATT) \cite{seo2017interpretable}}: Dual-ATT applies a local and a global attentions to encode the user(item) documents separately. The final representation is learned by concatenating representations learned from both local and global attention modules.

\textbf{A Zero-Attentional Relevance Matching Network for Review Modeling (ZARM) \cite{zeng2021zero}}: ZARM combines the concept of relevance matching and semantic matching, and uses an zero-attention schema to dynamically model user \& item representations.

\textbf{Attentive Aspect Modeling for Review-aware Recommendation (AARM) \cite{guan2019attentive}}: AARM is the state-of-the-art review-based model for top-N recommendation. It utilizes a Phrase-level Sentiment Analysis toolkit \cite{zhang2014explicit} to first extract aspect-opinion-sentiment triples from textual reviews. Then these triples are put into an attentive aspect-interaction module to learn the aspect-level interactions. The learned aspect-interaction vectors are concatenated with global-interaction latent factors learned from the implicit-feedback interaction matrix to compute the final affinity score. Note that AARM is intrinsically different from other text-as-feature models because it extracts information from review text only as additional features to be combined with a interaction-based model (e.g., a MF model). %of the existence of the global-interaction latent factors.

% \noindent \textbf{MPCN \cite{tay2018multi}, AHN \cite{dong2020asymmetrical}, Dual-ATT \cite{seo2017interpretable}, ZARM \cite{zeng2021zero}} etc. apply more complicated structures above the embedding layer and we skip the specific introduction to their detailed model designs here given the restriction of space.

% \vspace{-1pt}
\section{Experiments}
\subsection{Dataset Description}
% \vspace{-10pt}
\noindent We use three categories of data from Amazon Product Review Dataset \cite{mcauley2013hidden}\footnote{Amazon Product Review: \url{http://jmcauley.ucsd.edu/data/amazon/links.html}}. Specifically, we want to investigate how text-based representation learning models perform in the cold-start scenario, so we include both 5-core and 0-core datasets. Here k-core means each user/item has at least k interactions in the dataset. A detailed statistics of the dataset is showed at Table \ref{tab:statistics}. We use a randomized user-level 7:3 split, namely, for each user, 70\% of her total transactions are used for training and the rest for testing. We don't have a separate validation set since the interaction matrix is already very sparse in these datasets. To avoid the review information leaking problem mentioned by Catherine et al. \cite{catherine2017transnets}, all the reviews in testset are not used in the training of the models.

\begin{table}[h]
\caption{The basic statistics of the datasets}
\resizebox{\columnwidth}{!}{
\begin{tabular}{l|llll} \hline \hline
Dataset & \#users   & \#items & \#interactions & \#density \\ \hline 
Beauty-0core & \multicolumn{1}{c}{1,210,271} & \multicolumn{1}{c}{249,274}  & \multicolumn{1}{c}{2,023,070}   & \multicolumn{1}{c}{6.71e-6 } \\
Beauty-5core & \multicolumn{1}{c}{22,363} & \multicolumn{1}{c}{12,191}  & \multicolumn{1}{c}{198,502}   & \multicolumn{1}{c}{7.82e-4 } \\
Tools \& Home-0core & \multicolumn{1}{c}{1,212,047} & \multicolumn{1}{c}{260,657}  & \multicolumn{1}{c}{1,926,047}   & \multicolumn{1}{c}{6.09e-6 } \\
Tools \& Home-5core  & \multicolumn{1}{c}{16,638} & \multicolumn{1}{c}{10,217}  & \multicolumn{1}{c}{134,476}   & \multicolumn{1}{c}{7.92e-4 } \\
Electronics-0core& \multicolumn{1}{c}{4,201,696} & \multicolumn{1}{c}{476,001}  & \multicolumn{1}{c}{7,824,482}   & \multicolumn{1}{c}{6.91e-6 } \\
Electronics-5core & \multicolumn{1}{c}{192,403} & \multicolumn{1}{c}{63,001}  & \multicolumn{1}{c}{1,689,188}   & \multicolumn{1}{c}{1.39e-4 } \\ \hline \hline 
\end{tabular}
}
\label{tab:statistics}
\end{table}

\subsection{Evaluation}
We evaluate the ranking performances of all models using \textit{Hit Rate} (HR) and \textit{normalized Discounted Cumulative Gain} (nDCG). HR intuitively measures whether the test item is in the recommendation list and nDCG accounts for the position of the hit by assigning higher scores to hits at the top of the recommendation list.
We report the HR@10 and nDCG@10 on both 0-core and 5-core datasets. 

Previous implementations \cite{he2017neural,koren2008factorization,ning2011slim,xu2020commerce} rank the ground truth items along with \textit{k} randomly sampled negative items; According to Krichene and Rendle \cite{krichene2020sampled}, this may lead to inconsistent results and may not be a fair comparison between algorithms. 
Yet, ranking all the items in the candidate items pool will lead to prohibitive computation cost for some review-based models.
To reach a balance between consistency and efficiency, we use a two-stage retrieval \& reranking strategy. 
In the retrieval stage, we use MF to retrieve top 1,000 items for each user, and in the reranking stage, for each user, we rerank these items along with the ground truth items. 
Through this two-stage strategy, we reduce the overall time complexity to ranking all items using complex text-based models.
We also avoid the extreme case that randomly sampled negative test items are very irrelevant and leads to inconsistent recommendation performance in evaluation.

\subsection{Implementation Details}

\subsubsection{Text Processing}\hfill

\noindent We remove stopwords from the reviews and maintain a vocabulary of 50K most frequent words from the training corpus. To better catch the semantic information from textual reviews, we use Google's Word2Vec\footnote{https://code.google.com/archive/p/word2vec/} 300-dimensional embeddings pretrained on 100 billion words from Google News \cite{mikolov2013distributed}. For BERT-Rep method, top 512 tokens of each user \& item document are used, and for other review-based methods except for AARM, top 1,000 tokens are used. For AARM, we use the Phrase-level Sentiment Analysis toolkit Sentires \footnote{Sentires: \url{https://github.com/evison/Sentires}} to extract the aspect-opinion-sentiment triples from textual reviews. 
Specifically, we use the Word2Vec to convert the aspects into word/phrase embeddings.
%Specifically, we don't pretrain the aspect embeddings for AARM and use the Word2Vec to convert the aspects into word/phrase embeddings.

\subsubsection{Implementation} \hfill

\noindent We implemented MF, GMF, DeepCoNN, MPCN, AHN, Dual-ATT, NARRE, ZARM using PyTorch \footnote{PyTorch: \url{https://pytorch.org/}}. For BPR-HFT and JRL, we used the implementation from the JRL Repo \footnote{https://github.com/evison/JRL} and modified accordingly. For AARM, we used the implementation from AARM Repo \footnote{https://github.com/XinyuGuan01/Attentive-Aspect-based-Recommendation-Model} and modified accordingly. Our implementations will be available \footnote{https://github.com/zhichaoxu-shufe/understanding-reviews}. 

\subsubsection{Parameters \& Hyperparameters} \hfill

\noindent We train our models using Adam \cite{kingma2014adam} and SGD \cite{robbins1951stochastic}. We search the learning rate between 1e-1 and 1e-4, L2-regularization between 1e-1 and 1e-4, number of negative samples between 2 and 10, CNN window size between 3 and 10, dropout rate between 0.1 and 0.8, latent factor size between 16 and 128. In all the models we set the default latent factor size to 64 unless mentioned specifically.
% \vspace{-1pt}
\section{Result \& Analysis}

\begin{table*}[h]
\caption{The ranking performance (Hit Rate, nDCG) measured in \%; We highlight the model with best performance. $\clubsuit$ indicates its improvement over models in other two categories is statistically significant at 0.01 level with Paired t-test}
\resizebox{\textwidth}{!}
{
\begin{tabular}{l|l|cccccccccccc} \hline \hline
\multicolumn{2}{l|}{Dataset} &\multicolumn{4}{c|}{Tools \& Home} &\multicolumn{4}{c|}{Beauty} &\multicolumn{4}{c}{Electronics} \\ \cline{3-14} 
\multicolumn{2}{l|}{} &\multicolumn{2}{c|}{0-core} &\multicolumn{2}{c|}{5-core} &\multicolumn{2}{c|}{0-core} &\multicolumn{2}{c|}{5-core} &\multicolumn{2}{c|}{0-core} &\multicolumn{2}{c}{5-core} \\ \hline
\multicolumn{2}{l|}{Metrics} &\multicolumn{1}{c|}{Hit} &\multicolumn{1}{c|}{nDCG} &\multicolumn{1}{c|}{Hit} &\multicolumn{1}{c|}{nDCG} &\multicolumn{1}{c|}{Hit} &\multicolumn{1}{c|}{nDCG} &\multicolumn{1}{c|}{Hit} &\multicolumn{1}{c|}{nDCG} &\multicolumn{1}{c|}{Hit} &\multicolumn{1}{c|}{nDCG} &\multicolumn{1}{c|}{Hit} & nDCG \\ \hline
interaction-based
    & BPR-MF & 1.72 & 0.83 & 7.12 & 2.35 & 2.05 & 0.93 & 11.33 & 4.21 & 2.02 & 0.51 & 4.56 & 1.43 \\
    & BPR-GMF & 1.74 & 0.89 & 7.04 & 2.38 & 2.01 & 0.94 & 11.01 & 4.37 & 2.00 & 0.52 & 4.71 & 1.49 \\
    \hline
text-as-regularizer
    & BPR-HFT & 1.82 & 0.88 & 8.42 & 2.84 & 2.06 & 0.97 & 11.54 & 4.65 & 2.08 & 0.58 & 4.89 & 1.62 \\
    & JRL & $1.93$ & $0.97$ & $\textbf{8.84}^{\clubsuit}$ & $\textbf{3.01}^{\clubsuit}$ & $2.22$ & $1.04$ & $\textbf{12.04}^{\clubsuit}$ & $\textbf{5.03}^{\clubsuit}$ & $\textbf{2.15}$ & $\textbf{0.60}^{\clubsuit}$ & $\textbf{5.23}^{\clubsuit}$ & $\textbf{1.91}^{\clubsuit}$ \\
    \hline
text-as-only-feature
    & BERT & - & - & 5.35 & 1.69 & - & - & 8.87 & 4.05 & - & - & 3.67 & 1.05 \\
    & DeepCoNN & 1.45 & 0.69 & 4.82 & 1.64 & 1.80 & 0.81 & 7.96 & 3.79 & 1.78 & 0.42 & 3.58 & 1.17 \\
    & MPCN & 1.39 & 0.71 & 6.18 & 1.95 & 1.84 & 0.86 & 8.78 & 3.99 & 1.81 & 0.39 & 4.15 & 1.22 \\
    & AHN & 1.45 & 0.73 & 6.13 & 1.92 & 1.86 & 0.86 & 8.92 & 4.05 & 1.79 & 0.41 & 4.21 & 1.25 \\
    & Dual-ATT & 1.51 & 0.71 & 6.47 & 2.12 & 1.84 & 0.85 & 9.01 & 4.02 & 1.81 & 0.41 & 4.20 & 1.24 \\
    & NARRE & 1.59 & 0.78 & 6.80 & 2.25 & 1.94 & 0.84 & 8.94 & 4.01 & 1.85 & 0.45 & 4.29 & 1.29 \\
    & ZARM & 1.60 & 0.81 & 6.84 & 2.28 & 1.99 & 0.91 & 8.99 & 4.10 & 1.87 & 0.44 & 4.32 & 1.33 \\ 
\hline
text-as-additional-feature
    & AARM & $\textbf{2.05}^{\clubsuit}$ & $\textbf{1.01}$ & 7.94 & 2.68 & $\textbf{2.31}$ & $\textbf{1.09}$ & 11.59 & 4.76 & 2.12 & 0.54 & 4.95 & 1.56 \\ 
    \hline 
    \hline
\end{tabular}
}
\label{tab:results}
\end{table*}

\begin{table}[h]
\caption{The statistics of the reranking inference speed, measured in seconds per entry, with batch size 512}
\resizebox{\columnwidth}{!}{
\begin{tabular}{l|lll} \hline \hline
Model\textbackslash{}Dataset & \multicolumn{1}{c}{Tools \& Home-5core} & \multicolumn{1}{c}{Beauty-5core} & \multicolumn{1}{c}{Electronics-5core} \\ \hline
BPR-MF & \multicolumn{1}{c}{0.004} & \multicolumn{1}{c}{0.004} & \multicolumn{1}{c}{0.004} \\ \hline
BPR-GMF & \multicolumn{1}{c}{0.012} & \multicolumn{1}{c}{0.012} & \multicolumn{1}{c}{0.012} \\ \hline
BPR-HFT & \multicolumn{1}{c}{-} & \multicolumn{1}{c}{-} & \multicolumn{1}{c}{-} \\ \hline
JRL & \multicolumn{1}{c}{0.005} & \multicolumn{1}{c}{0.005} & \multicolumn{1}{c}{0.005} \\ \hline
DeepCoNN & \multicolumn{1}{c}{0.079} & \multicolumn{1}{c}{0.082} & \multicolumn{1}{c}{0.081} \\ \hline
AARM & \multicolumn{1}{c}{0.203} & \multicolumn{1}{c}{0.162} & \multicolumn{1}{c}{0.312} \\ \hline
NARRE & \multicolumn{1}{c}{0.170} & \multicolumn{1}{c}{0.173} & \multicolumn{1}{c}{0.170} \\ \hline
MPCN & \multicolumn{1}{c}{0.255} & \multicolumn{1}{c}{0.257} & \multicolumn{1}{c}{0.255} \\ \hline
AHN & \multicolumn{1}{c}{0.330} & \multicolumn{1}{c}{0.332} & \multicolumn{1}{c}{0.330} \\ \hline
Dual-ATT & \multicolumn{1}{c}{0.276} & \multicolumn{1}{c}{0.278} & \multicolumn{1}{c}{0.275} \\ \hline
ZARM & \multicolumn{1}{c}{0.383} & \multicolumn{1}{c}{0.385} & \multicolumn{1}{c}{0.383} \\ \hline \hline
\end{tabular} }
\label{tab:inference_speed}
\end{table}

\subsection{Top-N Recommendation Performance}
\label{section:performance}
We show the ranking results in Table \ref{tab:results}. We do not include the result of BERT model on 0-core dataset as the encoding takes too much time; 
performance-wise, we find: 

Among interaction-based models, BPR-GMF achieves about the same performances as BPR-MF in all datasets; this is the same as what was reported by Rendle \cite{rendle2020neural}. We argue that on sparse datasets like Amazon Product Reviews, BPR-GMF does not achieve significant improvement over BPR-MF because the deep MLP structure in equation \ref{ref:BPR-GMF} can not perfectly catch the complex interaction signals due to the lack of historical interactions for training.

% Among text-as-feature models, we notice that ZARM and NARRE consistently outperform other models. 
% For example, in 5-core Electronics dataset, ZARM achieves 2.8\% and 7.3\% improvement in HR and nDCG compared with Dual-ATT. 
% The performance difference between NARRE and ZARM is not statistically significant, so we consider they deliver about the same performances. 
% Compared to complex structures such as attention mechanism used in MPCN, AHN and Dual-ATT, NARRE utilizes a simple review-level attention and proves to be both effective and computationally efficient. 

Among text-as-feature models, we notice that AARM consistently outperform other models. For example, in 5-core Beauty dataset, AARM achieves 28.9\% and 16.1\% improvement in HR and nDCG respectively, compared with ZARM. 
We consider this performance boost is given by the global-interaction latent factors trained from the implicit-feedback interaction matrix.
Aside from AARM, the performance difference between NARRE and ZARM is not statistically significant, so we consider they deliver about the same performance. 
Compared to complex structures such as attention mechanism used in MPCN, AHN and Dual-ATT, NARRE utilizes a simple review-level attention and proves to be both effective and computationally efficient. 

% In 5-core datasets, we notice interaction-based models outperform text-as-feature models except for AARM. 
% For example, in Electronics, BPR-MF achieves 5.5\% improvement on Hit Rate, and 7.5\% improvement on nDCG; 
% We also find text-as-regularizer models achieve even better performances than interaction-based models and text-as-features models. For example, in Electronics, JRL has 14.7\%, 33.5\% improvement over BPR-MF in HR \& nDCG respectively; and 21.1\%, 43.6\% over ZARM. Our findings are in accordance with Sachedeva's \cite{sachdeva2020useful} argument that reviews are more effective as regularizer rather than as feature.
% Note that in Sachdeva's experiments, NARRE outperforms interaction-based models in rating prediction. We argue this is because of the intrinsic difference between rating prediction and top-N recommendation. We further discuss this intrinsic difference in section \ref{section:ratingvsranking}.

In 5-core datasets, we notice AARM consistently outperforms interaction-based models. For example, in Electronics dataset, AARM achieves 5.1\% and 9.1\% improvement in HR and nDCG respectively, compared with BPR-GMF.
This observation shows the aspect-interaction module can effectively capture the aspect-level features of the users and the items. 
Other text-as-feature models fail to deliver as good performance as interaction-based models. 
We also notice that text-as-regularizer models achieve the best performances overall compared with text-as-feature models and interaction-based models. For example, in Electronics dataset, JRL is 5.6\% better in HR and 22.4\% better in nDCG compared with AARM. 
Our findings are in accordance with Sachedeva's \cite{sachdeva2020useful} argument that reviews are more effective as regularizer rather than as feature. 
Note that in Sachdeva's experiments, NARRE outperforms interaction-based models in rating prediction while in our experiment, NARRE fails to outperform two interaction-based models. We argue this is because of the intrinsic difference between rating prediction and top-N recommendation. We further discuss this intrinsic difference in section \ref{section:ratingvsranking}.

% Complex models suffer from overfitting problem and often do not perform well in cold-start scenario. 
% We notice that in all three 0-core datasets, text-as-regularizer models significantly outperform both interaction-based models and text-as-feature models, and interaction-based models outperform text-as-only-feature models. Among text-as-only-feature models, we find DeepCoNN performs relatively better in 0-core datasets than in 5-core datasets. For example, DeepCoNN outperforms MPCN \& AHN in Tools \& Home 0-core dataset, and achieves about the same performance as MPCN, AHN \& D-ATT in Electronics 0-core dataset. We argue that DeepCoNN's TextCNN structure gives it an edge over complex attention-based models and suffer less from overfitting.

Profiling cold-start users \& items is a challenging task for E-commerce recommendation. 
Complex models suffer from overfitting problem and often don't perform well in cold-start scenario.
We notice that in all three 0-core datasets, JRL achieves about the same performance as AARM, and significantly outperforms other text-as-feature models and interaction-based models.
We consider AARM's good performance is mainly from pre-extracted aspects which are much more effective in cold-start scenario.
Furthermore, we notice that other text-as-feature models don't deliver as good performance as interaction-based models. This indicates that without sufficient training data, complex text-as-feature models are not able to model the user preferences as well as item features.
Among other text-as-feature models, we find DeepCoNN performs relatively better in 0-core datasets than in 5-core datasets. 
For example, DeepCoNN outperforms MPCN \& AHN in Tools \& Home 0-core dataset, and achieves about the same performance as MPCN, AHN \& D-ATT in Electronics 0-core dataset. 
We argue that DeepCoNN's TextCNN structure gives it an edge over complex attention-based models and suffer less from overfitting.

\subsection{Inference Speed}
Real-time response is also necessary for online E-commerce recommendation.
We report the comparison of inference speed at Table \ref{tab:inference_speed}. 
We leave out the BPR-HFT here as its implementation is in C++ while all other models are implemented in Python. 
We find the inference speed of interaction-based and text-as-regularizer models are fast; and the inference speed of text-as-only-feature models are significantly slower. 
Such phenomenon can be expected since in the inference stage, text-as-regularizer models only need to find the corresponding user and item latent factors and compute the affinity score through a similarity function, and text-as-feature models are slow since complicated neural network structures such as TextCNN, attention are used. 
We also notice that AARM's inference speed will increase significantly when there are more aspects in the dataset. 
For example, there are in total 1,987 aspects in Tools \& Home dataset and 690 aspects in Beauty dataset, and AARM's inference speed in Tools \& Home is 25\% more than in Beauty.
Compared with the performance reported at Table \ref{tab:results}, we conclude that AARM reaches the better balance between performance and computational complexity among all text-as-feature models. 
Overall, text-as-regularizer models are more computationally efficient.

\subsection{Analysis}
\subsubsection{Rating prediction and top-N recommendation are intrinsically different tasks} \hfill \label{section:ratingvsranking}

\noindent We formulate the structure of text-as-feature model for rating prediction as:
\begin{equation}
    Rep_{u} = f_u(\Gamma_{u}), \space Rep_{i} = f_i(\Gamma_{i}), \space \hat{r}_{u,i} = F(Rep_u, Rep_i)
\end{equation}
Each dimension of the latent factor can be regarded as an aspect of the user/item profile, and the final function $F(\cdot, \cdot)$ can be regarded as calculating the similarity between the user factor and the item factor; it varies from a simple inner product to a complicated deep neural network structure.

Explicit feedback problem i.e. problem aims to predict the level of affinity usually assume that the interaction has already happened, e.g. the user has watched the movie or has purchased the item. 
On the other hand, implicit feedback problem aims to predict whether the interaction will happen. Specifically, top-N recommendation aims to generate a short ranklist of items that the user may interact with to cover as many ground-truth items as possible. 
In modern E-commerce applications, this is more important as better top-N recommendation performance will increase the profit. 
Instead of fitting the absolute value of user-item ratings, user's preferences among different items are more important. 
We argue the intrinsic difference between two different types of tasks makes text-as-feature models designed specifically for explicit feedback problem e.g. rating prediction not suitable for implicit feedback problem, e.g. top-N recommendation, and existing model designs can not be migrated directly.

\subsubsection{User/item latent representations built purely from textual reviews do not deliver good top-N recommendation performance} \hfill

\noindent We observe that all text-as-feature models except for AARM fail to outperform vanilla matrix factorization model.
We argue there may be four reasons that lead to text-as-feature models' underperformances: 
(1) the reviews are of low quality and do not include much useful information; 
(2) existing model structures are limited thus can not extract useful information from reviews; 
(3) in implicit feedback problem, there is a gap between user leaving a piece of textual review and her actual interaction decision, which means reviews do not necessarily reflect the user's actual preference. BERT is a state-of-the-art pretrained contextualized language model and is supposed to catch useful semantic information from texts, but our BERT-Rep baseline is still outperformed by both interaction-based and text-as-regularizer models by a large margin; 
(4) following previous points, we regard rating prediction as an explicit-feedback task and top-N recommendation as an implicit-feedback task. 
In textual reviews, users explicitly express their affinities to the items, so we argue that textual reviews are intrinsically more suitable to be used in rating prediction than in top-N recommendation. 
Many other factors such as the price of the item, the availability, etc need to be taken into consideration when making the interact-or-not decision. 

So far we come to conclusion that purely relying on features extracted from textual reviews to build user/item representations is not a good approach for top-N recommendation task. 
JRL and AARM combines textual reviews with interactions data and achieves the best performance among all the models. 
This observation indicates that the combination of these two sources of information can deliver better performance. 
However, the optimum model structure is yet to be determined.

% Complex structures like attention are likely to overfit in cold-start scenario due to the relative smaller training corpus. But larger corpus also significantly increase the time complexity of training text-based models. Our experiment shows that existing text-as-feature models designed for rating prediction are not suitable for ranking task. Also, the BERT-Rep baseline we use can be regarded as a well-tuned text-as-feature model but is still outperformed by interaction-based models. 
% BERT is a state-of-the-art pretrained language model, and the BERT-Rep baseline is still outperformed by both interaction-based and text-as-regularizer models by a large margin.

% \begin{align}
% \begin{split}
% Rep_{u} &= f (\Gamma_{u}) \\
% Rep_{i} &= f (\Gamma_{i}) \\
% \hat{r}_{u,i} &= F (Rep_u, Rep_i) \\
% \end{split}
% \end{align}
\vspace{-1pt}
\section{Conclusion \& Future Work}
\vspace{-1pt}
In this work, we focus on the discussion of the better model structures to utilize textual reviews for E-commerce top-N recommendation. We adapt some existing text-as-features rating prediction models for top-N recommendation task. We also include existing top-N recommendation models and draw comparison from both performance and inference speed.
We find that due to the intrinsic difference between two different tasks, existing text-as-feature rating prediction model designs are not suitable for E-commerce top-N recommendation. 
We further discuss the possible reasons behind text-as-feature models' underperformances.
We come to conclusion that among existing review-based models, those only using textual reviews as features fail to outperform vanilla interaction-based models, and combining textual reviews with historical interactions data can deliver better performances.
Overall, text-as-regularizer models seem to be better at utilizing textual review information, giving better performances without increasing inference time complexity. 
We also provide our implementation of multiple strong baselines, hoping to shed light to the reproducibility issue in current recommender system research community.

Our future work will be designing better and more efficient models to utilize textual reviews for implicit feedback top-N recommendation task.

\section{Acknowledgement}
This work was supported in part by the School of Computing, University of Utah and in part by NSF IIS-2007398. Any opinions, findings and conclusions or recommendations expressed in this material are those of the authors and do not necessarily reflect those of the sponsor.

% \newpage
\bibliographystyle{ACM-Reference-Format}
\balance
\bibliography{reference}

\end{document}